# Linear Index Coding via Semidefinite Programming

Eden Chlamtáč[*]    Ishay Haviv[†]

**Abstract**

In the *index coding* problem, introduced by Birk and Kol (INFOCOM, 1998), the goal is to broadcast an $n$ bit word to $n$ receivers (one bit per receiver), where the receivers have *side information* represented by a graph $G$. The objective is to minimize the length of a codeword sent to all receivers which allows each receiver to learn its bit. For *linear* index coding, the minimum possible length is known to be equal to a graph parameter called *minrank* (Bar-Yossef et al., FOCS, 2006).

We show a polynomial time algorithm that, given an $n$ vertex graph $G$ with minrank $k$, finds a linear index code for $G$ of length $\widetilde{O}(n^{f(k)})$, where $f(k)$ depends only on $k$. For example, for $k = 3$ we obtain $f(3) \approx 0.2574$. Our algorithm employs a semidefinite program (SDP) introduced by Karger, Motwani and Sudan (J. ACM, 1998) for graph coloring and its refined analysis due to Arora, Chlamtac and Charikar (STOC, 2006). Since the SDP we use is not a relaxation of the minimization problem we consider, a crucial component of our analysis is an *upper bound* on the objective value of the SDP in terms of the minrank.

At the heart of our analysis lies a combinatorial result which may be of independent interest. Namely, we show an exact expression for the maximum possible value of the Lovász $\vartheta$-function of a graph with minrank $k$. This yields a tight gap between two classical upper bounds on the Shannon capacity of a graph.

[*]The Blavatnik School of Computer Science, Tel Aviv University, Tel Aviv 69978, Israel. Research supported in part by an ERC Advanced grant.

[†]The Blavatnik School of Computer Science, Tel Aviv University, Tel Aviv 69978, Israel. Supported by the Adams Fellowship Program of the Israel Academy of Sciences and Humanities.

# 1 Introduction

In the *index coding* problem a sender wishes to broadcast a word $x \in \{0,1\}^n$ to $n$ receivers $R_1, \ldots, R_n$ in a way that enables the $i$th receiver $R_i$ to recover the $i$th bit $x_i$. Every receiver has prior side information on $x$ induced by the (undirected) side information graph $G$ on the vertex set $[n] = \{1, 2, \ldots, n\}$. That is, the $i$th receiver $R_i$ knows $x_j$ for every $j$ adjacent to $i$ in $G$. The goal is to encode $x$ using a code of minimum length so that every receiver $R_i$ is able to recover $x_i$ given the encoded message and the side information that it has on $x$ according to $G$. For example, if $G$ is a complete graph then a 1-bit message that consists of the XOR of the $x_i$'s enables every receiver to recover its bit using the bits of its neighbors and the encoded message.

The study of index coding was initiated by Birk and Kol [6] and further developed by Bar-Yossef, Birk, Jayram and Kol [5]. This research is motivated by applications, such as video on demand and wireless networking, in which a network transmits information to clients, and during the transmission every client misses some of the information. At this step, the clients have side information on the transmitted information, and the network is interested in minimizing the broadcast length in a way that enables the clients to decode their target (see, e.g., [32]). The study of index coding is also motivated by the more general problem of *network coding*, introduced by Ahlswede et al. [1]. It was shown in [14] that network coding instances can be efficiently reduced to index coding instances.

For a graph $G$ we denote by $\beta_1(G)$ the minimum length of an index code for $G$. This graph parameter is well-known to be related to several classical graph parameters. Indeed, for a graph $G$, $\beta_1(G)$ is bounded from below by $\alpha(G)$, the maximum size of an independent set in $G$, as follows from the fact that an independent set in $G$ corresponds to a set of receivers with no mutual information. On the other hand, $\beta_1(G)$ is bounded from above by $\chi(\overline{G})$, the chromatic number of $\overline{G}$. Indeed, given a coloring of $\overline{G}$ we can construct an index code, each of whose bits is the XOR of the bits corresponding to the vertices in a color class.

In this paper we focus on *linear* index coding schemes, i.e., coding schemes in which the encoding function is linear. Bar-Yossef et al. [5] proved that the minimum length of a linear index code for a graph $G$ equals the minimum rank over $\mathbb{F}_2$ of a matrix that has ones in the diagonal and zeros in the entries that correspond to non-edges (and arbitrary values from $\mathbb{F}_2$ in the other entries). This graph parameter is called *minrank* and is denoted by $\text{minrk}_2(G)$. Clearly, $\text{minrk}_2(G)$ bounds $\beta_1(G)$ from above. It was shown in [5] that this bound is tight for several graph families, including perfect graphs, odd holes (odd-length cycles of length at least 5) and odd anti-holes (complements of odd holes). These results raised the question whether the minrank parameter characterizes the minimum length of general index codes. This question was answered in the negative by Lubetzky and Stav [28], who showed that for any $\varepsilon > 0$ and a sufficiently large $n$ there is an $n$ vertex graph $G$ with $\beta_1(G) \leq n^\varepsilon$ and $\text{minrk}_2(G) \geq n^{1-\varepsilon}$ (see [3] for additional examples). The following theorem summarizes the bounds mentioned above.

**Theorem 1.1** ([18, 19, 5]). *For every graph $G$,*

$$\alpha(G) \leq \beta_1(G) \leq \text{minrk}_2(G) \leq \chi(\overline{G}).$$

The original motivation to study the minrank parameter came from research on the *Shannon capacity* of graphs. In [30], Shannon introduced the *Shannon capacity* of a graph $G$, denoted by $c(G)$ and defined as the limit $\lim_{k\to\infty} \sqrt[k]{\alpha(G^k)}$, where $G^k$ is what is known as the *k*-fold strong graph product of $G$ with itself. This graph parameter, studied in the context of information theory, measures the effective size of an alphabet in zero-error communication over a noisy channel represented by the graph. Unfortunately, the behavior of the Shannon capacity of graphs in general is far from being well understood. Even the Shannon capacity of very simple graphs, such as the cycle of length 7, is not known. Therefore, upper and lower bounds on the Shannon capacity of graphs are of interest. Haemers [19] defined the minrank parameter and proved that it gives an upper bound on $c(G)$. A more well-known and tractable upper bound on $c(G)$ is the one of Lovász [27], known as *Lovász $\vartheta$-function* (see [25]). Although for most graphs the $\vartheta$-function is a tighter upper bound than the minrank bound, Haemers [18] showed that there are graphs for which the minrank bound is tighter. For example, it is known that for every odd $k$ there is a graph $S_k$ with $\mathrm{minrk}_2(S_k) = k$ and $\vartheta(S_k) = 2^{\frac{k-1}{2}} + 1$ (see [19]).

In this work we study the connection between the minrank parameter and the Lovász $\vartheta$-function of a graph. Specifically, we obtain a tight and exact upper bound on the Lovász $\vartheta$-function of a graph with minrank $k$. This bound compares two classical upper bounds on the Shannon capacity of a graph. We note that this research direction was also suggested in work on index coding by Bar-Yossef et al. [5]. In addition, we initiate the study of algorithms for linear index coding for graphs with bounded minrank. That is, given a graph $G$ with $\mathrm{minrk}_2(G) = k$, where $k$ is a fixed constant, find a linear index code for $G$, where the objective is to minimize the code length. Our bound on the Lovász $\vartheta$-function lies at the heart of the analysis of our algorithms.

We start with some background and then review our results.

**Graph Coloring.** For an integer $q$ a graph $G$ is *q-colorable* if it is possible to assign a color from $\{1, \ldots, q\}$ to every vertex so that no edge is monochromatic. Such an assignment is called a *q-coloring*. The *chromatic number* $\chi(G)$ of $G$ is the smallest $q$ for which $G$ is $q$-colorable. It is well-known that the problem of deciding whether a graph is $q$-colorable is NP-complete for any $q \geq 3$ [16] and can be easily solved in polynomial time for $q \in \{1, 2\}$.

For the problem of deciding between $\chi(G) \leq q$ and $\chi(G) \geq Q$, Khot proved in [23] NP-hardness with $Q = q^{\frac{\log q}{25}}$ for any large enough constant $q$. The largest $Q$ for which deciding between $\chi(G) \leq 3$ and $\chi(G) \geq Q$ is known to be NP-hard is $Q = 5$ [22, 17]. However, Dinur, Mossel and Regev [12] proved NP-hardness for any constants $3 \leq q < Q$ under a certain complexity assumption, related to Khot's unique games conjecture [24]. Recently, Dinur and Shinkar [13] improved the analysis of [12] and showed that, whenever $q \geq 4$, a similar hardness result holds even for $Q = \log^c n$ for some $c > 0$ where $n$ stands for the number of vertices in the graph. In addition, it is known that it is NP-hard to approximate the chromatic number of an $n$ vertex graph to within a factor of $n^{1-\varepsilon}$ for any $\varepsilon > 0$ [23, 33].

On the other hand, there is a long line of research on (randomized) polynomial time algorithms for graphs with bounded chromatic number. These algorithms, given an $n$ vertex $q$-colorable graph, find a $Q$-coloring of $G$ where $Q = O(n^\delta)$ for some constant $\delta = \delta(q) > 0$. For example, for $q = 3$, a simple algorithm due to Wigderson [31] colors a 3-colorable graph using $O(n^{\frac{1}{2}})$ colors. It is interesting to note that Wigderson's algorithm works even if the input graph $G$ is not 3-colorable but has the weaker property that any subgraph induced by the neighbors of a vertex is 2-colorable. In a series of increasingly sophisticated combinatorial algorithms, Blum [7] improved the number of colors to $\widetilde{O}(n^{\frac{3}{8}})$.[1] Then, Karger, Motwani and Sudan [21] introduced an algorithm for this problem based on a semidefinite relaxation and improved the number of colors to $\widetilde{O}(n^{\frac{1}{4}})$. Combining the combinatorial approach of [7] and the semidefinite relaxation of [21], Blum and Karger [7, 8] improved it to $\widetilde{O}(n^{\frac{3}{14}})$. Recently, a sequence of improvements by Arora, Chlamtac and Charikar [4] and Chlamtac [10] reduced the number of colors to $\widetilde{O}(n^{0.2111})$ and $\widetilde{O}(n^{0.2072})$ respectively. The situation with coloring $q$-colorable graphs for $q \geq 4$ is similar. The best known algorithm, due to Halperin et al. [20], colors $n$ vertex $q$-colorable graphs using $\widetilde{O}(n^{\alpha_q})$ colors, where $0 < \alpha_q < 1$ is some constant depending on $q$. For example, $\alpha_4 = \frac{7}{19} \approx 0.368$ and $\alpha_5 = \frac{97}{207} \approx 0.469$.

A major ingredient in the above coloring algorithms is a semidefinite programming (SDP) relaxation of the chromatic number [21] called *vector chromatic number*, which we denote by $\chi_v$. As a relaxation of the chromatic number, the vector chromatic number satisfies $\chi_v(G) \leq \chi(G)$ for every graph $G$. The main tool in [21] is a randomized rounding algorithm that given a graph $G$ with $\chi_v(G) = \kappa$ finds a large independent set (whose cardinality is monotone decreasing in $\kappa$). Interestingly, it was proven in [21] that a tighter relaxation, called the *strict vector chromatic number* and denoted by $\chi_v^{(s)}$, is related to Lovász $\vartheta$-function and satisfies for every graph $G$,

$$\chi_v^{(s)}(G) = \vartheta(\overline{G}). \tag{1}$$

**The Complexity of the minrank Parameter.** Consider the problem of deciding whether a graph $G$ satisfies $\text{minrk}_2(G) = k$ where $k$ is a fixed constant. For $k \in \{1, 2\}$ the problem is easy. Indeed, $\text{minrk}_2(G) = 1$ holds only for the complete graph, while $\text{minrk}_2(G) = 2$ holds precisely for the complement of a (non-empty) bipartite graph [29]. For $k = 3$, Peeters [29] proved that the problem is NP-complete (even if the input graphs are planar) via a reduction from 3-colorability. Langberg and Sprintson [26] observed that every graph $G$ satisfies

$$\text{minrk}_2(G) \geq \log_2 \chi(\overline{G}), \tag{2}$$

and concluded (using the upper bound in Theorem 1.1 and the hardness result of [12]) that it is NP-hard to approximate the minrank of a given graph to within any constant, assuming the same variant of the unique games conjecture as in [12]. In fact, using the recent result of [13] one can obtain a corresponding hardness result even for an approximation factor of $\Omega(\log \log n)$, where $n$ is the number of vertices. We note that the hardness result of Langberg and Sprintson [26] is proven for additional problems, including vector linear index coding, non-linear index coding, and the network coding problem.

---

[1]The $\widetilde{O}$ and $\widetilde{\Omega}$ notations are used to hide factors which are poly-logarithmic in $n$.

## 1.1 Our Contribution

### 1.1.1 Algorithms for Linear Index Coding

In this work we study algorithms for linear index coding for graphs with bounded minrank (recall that minrank measures the minimum length of a linear index code). Our approach to the problem is to design an algorithm that given a graph $G$ with $\text{minrk}_2(G) = k$ finds a coloring of $\overline{G}$ of as few colors as possible. As we have mentioned, such a coloring yields a linear index code whose length is the number of used colors. In what follows, for simplicity of presentation, the roles of $G$ and $\overline{G}$ will be reversed.

In the following discussion let us consider the case of $k = 3$ (recall that for $k \in \{1, 2\}$ the problem is easy). Let $G$ be a graph satisfying $\text{minrk}_2(\overline{G}) = 3$. Our goal is to find a coloring of $G$ with few colors. One strategy would be to use the fact that such a graph has the property that any subgraph induced by the neighbors of a vertex is 2-colorable (see Lemma 2.8). As mentioned earlier, for such graphs Wigderson's algorithm [31] can find an $O(n^{\frac{1}{2}})$-coloring of $G$.

Another strategy is to find the largest possible chromatic number $q$ of a graph $G$ satisfying $\text{minrk}_2(\overline{G}) = 3$, and to apply an algorithm for coloring $q$-colorable graphs to $G$. For example, by Inequality (2), $\chi(G) \leq 8$, and therefore the algorithm of [20] for coloring 8-colorable graphs gives an $\widetilde{O}(n^{\alpha_8})$-coloring of $G$ where $\alpha_8 = \frac{175}{271} \approx 0.646$, and hence a linear index code of such length. However, this can be somewhat improved. Peeters defined in [29] a graph family $G_k$ (see Section 3) such that for any $k$, $G_k$ is the graph that has a maximum chromatic number among all the graphs whose complement has minrank $k$. That is, for any $k$,

$$\chi(G_k) = \max\{\chi(G) \mid \text{minrk}_2(\overline{G}) = k\}.$$

It turns out that $G_3$ is a graph on 28 vertices with chromatic number 4, and this enables us to use the algorithm of [20] for coloring 4-colorable graphs and to get a polynomial time algorithm that given a graph $G$ with $\text{minrk}_2(\overline{G}) = 3$ finds a linear index code for $\overline{G}$ of length $\widetilde{O}(n^{\alpha_4})$ where $\alpha_4 \approx 0.469$.

In order to improve the number of colors used in the above algorithm we need to improve the analysis of the coloring algorithms in a way that uses the fact that our graphs have $\text{minrk}_2(\overline{G}) = 3$ and not only $\chi(G) \leq 4$. As mentioned earlier, the performance of the best known coloring algorithms crucially depends on the vector chromatic number of the input graph. This suggests studying the maximum possible vector chromatic number of $G$ assuming $\text{minrk}_2(\overline{G}) = 3$. A bound strictly smaller than 4 might imply an improved algorithm for our problem. In this work we observe that the graph family $G_k$ of Peeters [29] satisfies, for any $k$,

$$\chi_v(G_k) = \max\{\chi_v(G) \mid \text{minrk}_2(\overline{G}) = k\}.$$

Using (1) and our result on the Lovász $\vartheta$-function (Theorem 1.4) we obtain an exact expression for $\chi_v(G_k)$.[2] For example, for $k = 3$, we show that $\chi_v(G_3) = 1 + \frac{3}{\sqrt{2}} \approx 3.1213 < 4$ (see Corollary 3.9) and use it to show that it is possible to efficiently color a graph $G$ satisfying $\text{minrk}_2(\overline{G}) = 3$ using fewer colors than guaranteed by the algorithm for coloring 4-colorable graphs of [20]. This, combined with additional properties of graphs with $\text{minrk}_2(\overline{G}) = 3$ and some techniques used in [4], yields the following theorem and its immediate corollary. We also prove similar results for general $k \geq 3$ (see Theorem 4.3).

---
[2]It turns out that the gap between $\chi(G_k)$ and $\chi_v(G_k)$ is exponentially large in $k$ (see Section 5 for details).

**Theorem 1.2.** *There exists a randomized polynomial time algorithm that given an n vertex graph G with* $\text{minrk}_2(\overline{G}) = 3$ *finds an* $\widetilde{O}(n^{0.2574})$*-coloring of G.*

**Corollary 1.3.** *There exists a randomized polynomial time algorithm that given an n vertex graph G, for which there is a linear index code of length* 3*, finds a linear index code for G of length* $\widetilde{O}(n^{0.2574})$.

One obstacle presented by our approach is that the SDP relaxation we use (the vector chromatic number of the complement graph) is not a relaxation of the minrank parameter. This is in contrast to the usual framework (say, for combinatorial minimization problems), in which the SDP is a relaxation of the minimization problem at hand, which then automatically bounds the SDP optimum from above by the actual optimum (of the original optimization problem). The fact that we deviate from this approach necessitates an additional crucial component of the analysis, namely, bounding the largest possible gap between the vector chromatic number and the minrank parameter.

It is then natural to ask whether an improved performance guarantee can be achieved by a more straightforward approach. While the question of further improvements remains open, we do note that using an SDP relaxation for minrank based on the structure of the graph $G_k$, we can achieve the same performance guarantee by transforming the solution of the new SDP into a vector $\kappa$-coloring of the complement graph, where $\kappa = \chi_v(G_k)$ (see the discussion in Section 5).

#### 1.1.2 Minrank versus Lovász $\vartheta$-function

In addition to our algorithmic application, the graph family $G_k$ is interesting from a combinatorial point of view. Recall that both $\vartheta(G)$ and $\text{minrk}_2(G)$ are upper bounds on the Shannon capacity $c(G)$ of a graph. A natural question to ask is how large $\vartheta(G)$ can be if G satisfies $\text{minrk}_2(G) = k$, or, equivalently, how bad the $\vartheta(G)$ upper bound can be, compared to the bound $\text{minrk}_2(G)$. We show that the largest $\vartheta(G)$ for a graph G with $\text{minrk}_2(G) = k$ is attained at $\overline{G_k}$ for which we calculate the exact $\vartheta$ value.

**Theorem 1.4.** *For every k, every graph G with* $\text{minrk}_2(G) = k$ *satisfies*

$$\vartheta(G) \leq 2^{\frac{k}{2}} + 1 - 2^{1-\frac{k}{2}}.$$

*In addition, equality holds for the graph* $\overline{G_k}$.

Our calculation of the $\vartheta$ value of $\overline{G_k}$ is based on strong symmetry properties of $G_k$ which relate the $\vartheta$ value to the spectrum of $G_k$. To calculate the spectrum, we fix a vertex $v$ and define a partition of the vertex set into 5 equivalence classes with the following property: two vertices $u_1$ and $u_2$ are in the same class if and only if there exists an automorphism $f$ such that $f(v) = v$ and $f(u_1) = u_2$. In group theoretic terms, these are the orbits of the vertex stabilizer of $v$ relative to the automorphism group of the graph.

We show that the the eigenvalues of $G_k$ coincide with the eigenvalues of a weighted graph obtained by contracting each equivalence class to a single vertex. This reduces the eigenvalue calculation to finding the spectrum of a $5 \times 5$ matrix. For details see Section 3.3.

## 1.2 Outline

In the next section we provide the background and definitions needed throughout the paper, including two equivalent definitions of the minrank parameter. In Section 3 we define the graph family $G_k$, prove its properties which are needed for the analysis of our algorithms, and prove Theorem 1.4. Finally, in Section 4 we present and analyze our algorithms for linear index coding and prove Theorem 1.2 and its extension to general $k$. We end with a discussion in Section 5.

## 2 Preliminaries

### 2.1 Index Coding

In the index coding problem a sender wishes to broadcast a word $x \in \mathbb{F}_2^n$ to $n$ receivers $R_1, \ldots, R_n$. Every receiver $R_i$ knows some fixed subset of the bits of $x$ and is interested solely in $x_i$. An *index code* for this setting is a code over $\mathbb{F}_2$, which enables $R_i$ to recover $x_i$ for every $x \in \mathbb{F}_2^n$ and $i \in [n]$.

The index coding problem can be stated as a graph parameter. For a directed graph $G$ and a vertex $v$ let $\Gamma_G^+(v)$ denote the set of out-neighbors of $v$ in $G$, and for $x \in \mathbb{F}_2^n$ and $S \subseteq [n]$ let $x|_S$ denote the restriction of $x$ to the coordinates of $S$. The setting of the definition of an index code is characterized by the directed *side information graph* $G$ on the vertex set $[n]$ where $(i, j)$ is an edge if and only if the receiver $R_i$ knows $x_j$. An index code of length $\ell$ for $G$ (over $\mathbb{F}_2$) is a function $E : \mathbb{F}_2^n \to \mathbb{F}_2^\ell$ and functions $D_1, \ldots, D_n$, so that for all $i \in [n]$ and $x \in \mathbb{F}_2^n$, $D_i(E(x), x|_{\Gamma_G^+(i)}) = x_i$. The definition of an index code is naturally extended to undirected graphs by replacing every undirected edge by two oppositely directed edges, and this is the focus of the current work. We say that the index code is *linear* if the encoding function $E$ is linear.

### 2.2 Vector Chromatic Number

The vector chromatic number and the strict vector chromatic number [21] are defined as follows.

**Definition 2.1.** *For a graph $G = (V, E)$ the* vector chromatic number *of $G$, denoted by $\chi_v(G)$, is the minimal real value of $\kappa$ such that there exists an assignment of a unit vector $w_i$ to each vertex $i$ satisfying the inequality $\langle w_i, w_j \rangle \leq -\frac{1}{\kappa-1}$ whenever $i$ and $j$ are adjacent in $G$. Such an assignment is a vector $\kappa$-coloring of $G$.*

**Definition 2.2.** *For a graph $G = (V, E)$ the* strict vector chromatic number *of $G$, denoted by $\chi_v^{(s)}(G)$, is the minimal real value of $\kappa$ such that there exists an assignment of a unit vector $w_i$ to each vertex $i$ satisfying the inequality $\langle w_i, w_j \rangle = -\frac{1}{\kappa-1}$ whenever $i$ and $j$ are adjacent in $G$. Such an assignment is a strict vector $\kappa$-coloring of $G$.*

Observe that the (strict) vector chromatic number is a relaxation of the chromatic number. Moreover, for every graph G,
$$\chi_v(G) \leq \chi_v^{(s)}(G) \leq \chi(G).$$
There are, however, graphs which are vector k-colorable but are not k-colorable. It is known that a (strict) vector $\kappa$-coloring of a graph G, if exists, can be found in polynomial time by solving a semidefinite program (see [21] for details). We note that the Lovász $\vartheta$-function, introduced in [27], is known to satisfy $\chi_v^{(s)}(G) = \vartheta(\overline{G})$ for every graph G [21].

The following lemma was used in the analysis of the coloring algorithm in [21]. For a graph G and a vertex v in G we use $\Gamma_G(v)$ to denote the set of neighbors of v in G. If G is clear from the context we omit the subscript. For a subset S of the vertices, $G[S]$ denotes the subgraph of G induced by the vertices in S.

**Lemma 2.3** ([21]). *Let $G = (V, E)$ be a graph satisfying $\chi_v(G) = \kappa$ for some $\kappa > 2$. Then, for every vertex $v \in V$, the subgraph of G induced by the neighbors of v satisfies $\chi_v(G[\Gamma(v)]) \leq \kappa - 1$.*

## 2.3 Minrank

The minrank of a graph (over $\mathbb{F}_2$) is defined as follows.

**Definition 2.4.** *Let $A = (a_{ij})$ be an n by n matrix over $\mathbb{F}_2$. We say that A represents an n vertex graph G over $\mathbb{F}_2$ if $a_{ii} = 1$ for all i, and $a_{ij} = 0$ whenever i and j are distinct non-adjacent vertices in G. The minrank of a graph G over $\mathbb{F}_2$ is defined as*
$$\mathrm{minrk}_2(G) = \min\{\mathrm{rank}_{\mathbb{F}_2}(A) \mid A \text{ represents } G \text{ over } \mathbb{F}_2\}.$$

Bar-Yossef et al. [5] proved that the minrank parameter characterizes the minimum length of a linear index code, as stated below.

**Theorem 2.5** ([5]). *For every graph G, the minimum length of a linear index code for G over $\mathbb{F}_2$ equals $\mathrm{minrk}_2(G)$.*

Throughout the paper we need an equivalent characterization of minrank due to Peeters [29]. In what follows we define an orthogonal bi-representation over $\mathbb{F}_2$ of a graph and then use it to characterize the minrank parameter in Lemma 2.7

**Definition 2.6.** *For a graph $G = (V, E)$ an orthogonal bi-representation of G in $\mathbb{F}_2^k$ is an assignment of two vectors $a_v^1$ and $a_v^2$ in $\mathbb{F}_2^k$ to each vertex v of G, such that*

1. $\langle a_v^1, a_v^2 \rangle = 1$ *for every $v \in V$ and*
2. $\langle a_u^1, a_v^2 \rangle = \langle a_v^1, a_u^2 \rangle = 0$ *for every two distinct non-adjacent vertices $u, v \in V$.*

**Lemma 2.7** ([29]). *For every graph G, $\mathrm{minrk}_2(G)$ is the smallest k for which there exists an orthogonal bi-representation of G in $\mathbb{F}_2^k$.*

We note that Lemma 2.7 follows from Definitions 2.4 and 2.6 and the linear algebra fact that the rank of an $n$ by $n$ matrix $M$ is the smallest $k$ for which $M = A \cdot B$ for some two matrices $A$ and $B$ of dimensions $n \times k$ and $k \times n$ respectively.

The following lemma is an analogue of Lemma 2.3 for minrank and is used in the analysis of our algorithms.

**Lemma 2.8.** *Let $G = (V, E)$ be a graph satisfying $\mathrm{minrk}_2(\overline{G}) = k$ for some $k \geq 2$. Then, for every vertex $v \in V$, the subgraph of $G$ induced by the neighbors of $v$ satisfies $\mathrm{minrk}_2(\overline{G[\Gamma_G(v)]}) \leq k - 1$.*

**Proof:** Let $G$ be a graph that satisfies $\mathrm{minrk}_2(\overline{G}) = k$ and let $v$ be a vertex in $G$. Observe that any matrix $M$ that represents $\overline{G[\Gamma_G(v) \cup \{v\}]}$ satisfies $M_{v,v} = 1$ and $M_{u,v} = M_{v,u} = 0$ for any $u \in \Gamma_G(v)$. This implies that $\mathrm{minrk}_2(\overline{G[\Gamma_G(v) \cup \{v\}]}) = 1 + \mathrm{minrk}_2(\overline{G[\Gamma_G(v)]})$, which yields that

$$\mathrm{minrk}_2(\overline{G[\Gamma_G(v)]}) = \mathrm{minrk}_2(\overline{G[\Gamma_G(v) \cup \{v\}]}) - 1 \leq \mathrm{minrk}_2(\overline{G}) - 1 = k - 1.$$

∎

## 2.4 Graph Symmetry

In the following we define vertex-transitive and edge-transitive graphs and state some of their properties (for proofs, see [25]). We let $\lambda_1(G)$ and $\lambda_n(G)$ denote the largest and smallest eigenvalues of an $n$ vertex graph $G$.

**Definition 2.9.** *An* automorphism *of a graph $G = (V, E)$ is a permutation $f : V \to V$ that preserves edges, i.e., for every $u, v \in V$ it holds that $\{u, v\} \in E$ if and only if $\{f(u), f(v)\} \in E$. The graph $G$ is* vertex-transitive *if for every $u, v \in V$ there exists an automorphism $f$ of $G$ that satisfies $f(u) = v$. The graph $G$ is* edge-transitive *if for every two edges $\{u, v\}, \{x, y\} \in E$ there exists an automorphism $f$ of $G$ that satisfies $\{f(u), f(v)\} = \{x, y\}$.*

**Lemma 2.10.** *If $G$ is a vertex-transitive graph on $n$ vertices then $\vartheta(G) \cdot \vartheta(\overline{G}) = n$.*

**Lemma 2.11.** *If $G$ is an edge-transitive graph on $n$ vertices then $\vartheta(G) = n \cdot \frac{-\lambda_n(G)}{\lambda_1(G) - \lambda_n(G)}$.*

Combining Lemmas 2.10 and 2.11 we obtain the following corollary.

**Corollary 2.12.** *If $G$ is a vertex-transitive and edge-transitive graph on $n$ vertices then $\vartheta(\overline{G}) = 1 - \frac{\lambda_1(G)}{\lambda_n(G)}$.*

We also need the following straightforward lemma.

**Lemma 2.13.** *If $G$ is an edge-transitive graph then $\chi_v(G) = \chi_v^{(s)}(G)$.*

**Proof:** It suffices to show that a vector $\kappa$-coloring of a graph $G = (V, E)$ can be transformed into a strict vector $\kappa$-coloring of $G$. Let $(w_i)_{i \in V}$ be a vector $\kappa$-coloring of $G$ and let $\mathrm{Aut}(G)$ denote the group of all the automorphisms of $G$. For every $i \in V$ let $z_i$ be the normalized concatenation of all the $w_{f(i)}$ for $f \in \mathrm{Aut}(G)$. Observe that $(z_i)_{i \in V}$ is also a vector $\kappa$-coloring of $G$ and that

$$\langle z_i, z_j \rangle = \frac{1}{|\mathrm{Aut}(G)|} \cdot \sum_{f \in \mathrm{Aut}(G)} \langle w_{f(i)}, w_{f(j)} \rangle.$$

Since Aut($G$) acts transitively on the edge set $E$, it can be seen that the number of automorphisms of $G$ mapping $e$ to $e'$ is equal for every two edges $e$ and $e'$ (see Claim 3.15). This implies that for every edge $\{i, j\}$,

$$\frac{1}{|\mathrm{Aut}(G)|} \cdot \sum_{f \in \mathrm{Aut}(G)} \langle w_{f(i)}, w_{f(j)} \rangle = \frac{1}{|E|} \cdot \sum_{\{i',j'\} \in E} \langle w_{i'}, w_{j'} \rangle,$$

and hence all the inner products $\langle z_i, z_j \rangle$ for edges $\{i, j\}$ are equal, and we are done. ∎

## 3 The Graph Family $G_k$

In this section we define the graph family $G_k$ introduced in [29] and prove several properties that it has which are crucial for the analysis of our algorithm for linear index coding. We obtain a tight and exact upper bound on the Lovász $\vartheta$-function of graphs with minrank at most $k$ and prove Theorem 1.4. The inner products in the definition of $G_k$ are over $\mathbb{F}_2$. We denote by $e_i$ the vector in $\mathbb{F}_2^k$ that has a nonzero entry only in the $i$th coordinate.

For a natural $k \geq 1$ we define the undirected graph $G_k = (V, E)$ whose vertex set is

$$V = \{(v_1, v_2) \in \mathbb{F}_2^k \times \mathbb{F}_2^k \mid \langle v_1, v_2 \rangle = 1\},$$

and two vertices $(u_1, u_2)$ and $(v_1, v_2)$ are adjacent if and only if $\langle u_1, v_2 \rangle = \langle v_1, u_2 \rangle = 0$. Observe that $|V| = (2^k - 1) \cdot 2^{k-1}$ and that $G_k$ is regular with degree $(2^{k-1} - 1) \cdot 2^{k-2}$.

**Claim 3.1.** $\alpha(\overline{G_k}) = c(\overline{G_k}) = \mathrm{minrk}_2(\overline{G_k}) = k$.

**Proof:** By the definition of $G_k$, it is clear that there exists an orthogonal bi-representation of $\overline{G_k}$ in $\mathbb{F}_2^k$ (identify every vertex with the vectors in its name), and thus $\mathrm{minrk}_2(\overline{G_k}) \leq k$. On the other hand, the set of vertices $\{(e_i, e_i)\}_{i=1}^k$ is an independent set in $\overline{G_k}$. Since the size of an independent set in a graph bounds from below its Shannon capacity and since the minrank bounds it from above, we obtain $k \leq \alpha(\overline{G_k}) \leq c(\overline{G_k}) \leq \mathrm{minrk}_2(\overline{G_k}) \leq k$, and we are done. ∎

Observe that Lemma 2.7 implies that the minrank of a graph $G$ is precisely the smallest $k$ for which there exists a homomorphism from $\overline{G}$ to $G_k$. Peeters [29] used this observation to show that $G_k$ has maximum chromatic number among the graphs whose complement has minrank $k$. Similarly, this also holds for the (strict) vector chromatic number.

**Lemma 3.2 ([29]).** $\chi(G_k) = \max\{\chi(G) \mid \mathrm{minrk}_2(\overline{G}) = k\}$.

**Lemma 3.3.** $\chi_v^{(s)}(G_k) = \max\{\chi_v^{(s)}(G) \mid \mathrm{minrk}_2(\overline{G}) = k\}$.

**Proof:** By Claim 3.1, $\mathrm{minrk}_2(\overline{G_k}) = k$, and hence $\chi_v^{(s)}(G_k) \leq \max\{\chi_v^{(s)}(G) \mid \mathrm{minrk}_2(\overline{G}) = k\}$. For the other direction, denote $\kappa = \chi_v^{(s)}(G_k)$ and fix a strict vector $\kappa$-coloring $\{w_i\}$ of $G_k$. Let $G$ be a graph that satisfies $\mathrm{minrk}_2(\overline{G}) = k$ and let $h$ be a homomorphism from $G$ to $G_k$. To every vertex $x$ of $G$ assign the vector $w_{h(x)}$, and observe that this is a vector $\kappa$-coloring of $G$. ∎

The following corollary says that $\overline{G_k}$ attains the maximum Lovász $\vartheta$-function among the graphs with minrank $k$. It follows immediately from the previous lemma and Equation (1).

**Corollary 3.4.** $\vartheta(\overline{G_k}) = \max\{\vartheta(\overline{G}) \mid \mathrm{minrk}_2(G) = k\}$.

## 3.1 Symmetry Properties of $G_k$

We turn to prove that $G_k$ is vertex-transitive and edge-transitive. By Corollary 2.12, this implies that the strict vector chromatic of $G_k$ can be expressed in terms of the smallest and largest eigenvalues of $G_k$. We start with the following lemma. Recall that $e_i$ stands for the vector in $\mathbb{F}_2^k$ that has a nonzero entry only in the $i$th coordinate.

**Lemma 3.5.** *For every vertex $v = (v_1, v_2)$ in $G_k$ satisfying $(v_1)_1 = (v_2)_1 = 1$ there exists an automorphism $f$ of $G_k$ such that $f(e_1, e_1) = v$. In addition, if $v$ is adjacent to $(e_2, e_2)$ in $G_k$, the automorphism $f$ satisfies $f(e_2, e_2) = (e_2, e_2)$.*

**Proof:** Let $A \in \mathbb{F}_2^{k \times k}$ be the matrix whose first column is the vector $v_1$, its first row is the vector $v_2$ (this is possible since $(v_1)_1 = (v_2)_1 = 1$), and the remaining $(k-1) \times (k-1)$ block consists of the identity matrix of order $k-1$. Observe that $Ae_1 = v_1$ and that $A^t v_2 = e_1$. We claim that $A$ is invertible. To see it, notice that its last $k-1$ rows are linearly independent, so it is enough to show that the vector $v_2$ is not a linear combination of them. This is true because such a linear combination must have the last $k-1$ entries of $v_2$ as coefficients, but this will give us a vector that differs from $v_2$ in its first entry since $\langle v_1, v_2 \rangle = 1$.

Observe that any invertible matrix $A \in \mathbb{F}_2^{k \times k}$ satisfies $\langle x, y \rangle = \langle Ax, A^{-t}y \rangle$ for every two vectors $x, y \in \mathbb{F}_2^k$. This implies that the function $f$ that maps any vertex $(x, y) \in V$ to $(Ax, A^{-t}y)$ is an automorphism of $G_k$. In addition, $f$ satisfies $f(e_1, e_1) = (Ae_1, A^{-t}e_1) = (v_1, v_2) = v$. Finally, notice that if $v$ is adjacent to $(e_2, e_2)$ then $(v_1)_2 = (v_2)_2 = 0$, and this implies that $f(e_2, e_2) = (e_2, e_2)$, as required. ∎

**Lemma 3.6.** *The graph $G_k$ is vertex-transitive.*

**Proof:** Consider a vertex $v = (v_1, v_2)$ in $V$. It suffices to show that there exists an automorphism of $G_k$ that maps $v$ to $(e_1, e_1)$. Observe that there exists an $i$ for which $(v_1)_i = (v_2)_i = 1$, and let $f_1$ denote the automorphism of $G_k$ that switches the first and the $i$th coordinate in the two vectors of each vertex. By Lemma 3.5, there exists an automorphism $f_2$ of $G_k$ that maps $(e_1, e_1)$ to $f_1(v)$, which implies that $f_2^{-1} \circ f_1$ is an automorphism as desired. ∎

**Lemma 3.7.** *The graph $G_k$ is edge-transitive.*

**Proof:** Let $\{u, v\}$ be an edge in $G_k$. It suffices to show that there exists an automorphism of $G_k$ that maps $u$ to $(e_1, e_1)$ and $v$ to $(e_2, e_2)$. Since $G_k$ is vertex-transitive there exists an automorphism $f_1$ of $G_k$ that maps $v$ to $(e_2, e_2)$. Notice that $f_1(u)$ and $(e_2, e_2)$ are adjacent in $G_k$, and thus the second entry of the two vectors of $f_1(u)$ is zero. Hence, there exists an $i \neq 2$ for which the two vectors of the vertex $f_1(u)$ are nonzero in the $i$th entry. Denote by $f_2$ the automorphism of $G_k$ that switches the first and the $i$th coordinate in the two vectors of each vertex, and observe that $f_2(f_1(u))$ is adjacent to $(e_2, e_2)$. By Lemma 3.5, there exists an automorphism $f_3$ of $G_k$ that maps $(e_1, e_1)$ to $f_2(f_1(u))$ and $(e_2, e_2)$ to itself. Observe that $g = f_3^{-1} \circ f_2 \circ f_1$ is an automorphism of $G_k$ that satisfies $g(u) = (e_1, e_1)$ and $g(v) = (e_2, e_2)$. ∎

## 3.2 Minrank versus Lovász $\vartheta$-function

Now we are ready to derive our tight bound on the $\vartheta$-function of graphs with minrank $k$. Combining Lemma 2.13, Equation (1) and Corollary 2.12 we obtain that

$$\chi_v(G_k) = \chi_v^{(s)}(G_k) = \vartheta(\overline{G_k}) = 1 - \frac{\lambda_1(G_k)}{\lambda_n(G_k)},$$

where $n$ denotes the number of vertices in $G_k$. Recall that $G_k$ is regular and hence $\lambda_1(G_k)$ equals the degree of its vertices, i.e., $(2^{k-1} - 1) \cdot 2^{k-2}$. We need the following lemma proven in Section 3.3, which immediately implies the corollary that follows it.

**Lemma 3.8.** *For every $k$, $\lambda_n(G_k) = -2^{\frac{3k}{2}-3}$.*

**Corollary 3.9.** *For every $k$, $\chi_v(G_k) = \chi_v^{(s)}(G_k) = \vartheta(\overline{G_k}) = 2^{\frac{k}{2}} + 1 - 2^{1-\frac{k}{2}}$.*

The following theorem follows from the above discussion using Corollary 3.4.

**Theorem 3.10.** *For every $k$, every graph $G$ with $\mathrm{minrk}_2(G) \leq k$ satisfies*

$$\chi_v(\overline{G}) \leq \chi_v^{(s)}(\overline{G}) = \vartheta(G) \leq 2^{\frac{k}{2}} + 1 - 2^{1-\frac{k}{2}}.$$

*In addition, equalities hold for the graph $\overline{G_k}$.*

## 3.3 Spectral Analysis of $G_k$

In what follows we show how graph symmetries allow us to reduce the problem of eigenvalue calculation to identifying a certain partition of the vertex set. We then identify this partition for the graph $G_k$, thus concluding its spectral analysis.

### 3.3.1 Spectral Analysis via Graph Symmetry

Let us start by reviewing the following basic group theoretic definitions.

**Definition 3.11.** *Let $\mathcal{G}$ be a group which acts on a set $X$. For a group element $g \in \mathcal{G}$, denote its action on an element $x \in X$ by $g(x)$. For any element $x \in X$, its* stabilizer *(relative to $\mathcal{G}$), is the subgroup*

$$\mathrm{Stab}_\mathcal{G}(x) = \{g \in \mathcal{G} \mid g(x) = x\}.$$

**Definition 3.12.** *For a group $\mathcal{G}$ which acts on a set $X$, the* orbit *of an element $x \in X$ is the set*

$$\mathrm{Orb}_\mathcal{G}(x) = \{g(x) \mid g \in \mathcal{G}\}.$$

**Remark 3.13.** *The orbits of a set $X$ form a partition of $X$. That is, for any two elements $x, y \in X$, the orbits $\mathrm{Orb}_\mathcal{G}(x)$ and $\mathrm{Orb}_\mathcal{G}(y)$ are either equal or disjoint.*

**Definition 3.14.** *A group $\mathcal{G}$ acts* transitively *on a set $X$ if for all $x, y \in X$ there is a group element $g \in \mathcal{G}$ such that $g(x) = y$.*

**Claim 3.15.** *Let $\mathcal{G}$ be a group that acts transitively on a set X. Then the sets $T_{x,y} = \{g \in \mathcal{G} \mid g(x) = y\}$ have the same cardinality for every $x, y \in X$.*

**Proof:** Note that by transitivity, for every $x, y \in X$ the set $T_{x,y}$ is nonempty. Moreover, the set $T_{x,y}$ is a coset of the stabilizer group $T_{x,x}$. Therefore, for a fixed $x \in X$ all the sets in $\{T_{x,y}\}_{y \in X}$ have the same cardinality. By transitivity, this cardinality does not depend on $x$. ∎

We will be interested in the orbits of the vertex set $V(G_k)$ relative to the stabilizer subgroup of a fixed vertex $v_0$. That is, the orbits induced by the group of automorphisms of $G_k$ which map $v_0$ to itself. To simplify notation, for any graph $G = (V, E)$ and any vertex $v \in V$, we will denote

$$\text{AutStab}_G(v) = \text{Stab}_{\text{Aut}(G)}(v),$$
$$\text{StabOrb}_G(v) = \{\text{Orb}_{\text{AutStab}_G(v)}(u) \mid u \in V\}.$$

This partition will be crucial in finding the eigenvalues of $G_k$, as it will essentially allow us to treat each orbit as an individual vertex. This is done via the following lemma, which shows that for every eigenvalue there is an eigenvector with a structure that depends on the above partition.

**Lemma 3.16.** *Let $G = (V, E)$ be a vertex-transitive graph, let $\lambda$ be an eigenvalue of its adjacency matrix $A_G$, and let $v \in V$ be any vertex in the graph. Then, there exists a (nonzero) eigenvector $a_\lambda$ of $A_G$ corresponding to $\lambda$ such that for every orbit $O \in \text{StabOrb}_G(v)$ all the values of $a_\lambda$ in the coordinates corresponding to O are equal.*

**Proof:** Let $b_0$ be an eigenvector of $A_G$ corresponding to the eigenvalue $\lambda$. It must have at least one nonzero coordinate. Note that permuting the coordinates of an eigenvector by an automorphism of $G$ always yields an eigenvector corresponding to the same eigenvalue. Thus, by the vertex-transitivity of $G$, there is an automorphism which, applied to $b_0$, yields an eigenvector $b_1$ corresponding to the same eigenvalue, such that $(b_1)_v \neq 0$. Now, consider the effect of applying any automorphism in the stabilizer $\text{AutStab}_G(v)$ to $b_1$. This does not change the value in the coordinate corresponding to $v$. Moreover, any linear combination of such eigenvectors is also an eigenvector corresponding to $\lambda$, since all such eigenvectors lie in the same eigenspace. Define

$$a_\lambda = \frac{1}{|\text{AutStab}_G(v)|} \cdot \sum_{f \in \text{AutStab}_G(v)} f(b_1),$$

and observe, using Claim 3.15, that for every orbit $O$ and vertex $u \in O$,

$$(a_\lambda)_u = \frac{1}{|\text{AutStab}_G(v)|} \cdot \sum_{f \in \text{AutStab}_G(v)} (f(b_1))_u = \frac{1}{|O|} \cdot \sum_{u' \in O} (f(b_1))_{u'}.$$

∎

**Corollary 3.17.** *For every vertex-transitive graph $G = (V, E)$ and a vertex $v \in V$, the adjacency matrix $A_G$ has the same set of eigenvalues as the matrix M indexed by the orbits in $\text{StabOrb}_G(v)$ defined as follows: $M_{O_1, O_2}$ is the number of edges between $O_1$ and $O_2$ normalized by the number of vertices in $O_1$, and $M_{O_1, O_1}$ is twice the number of edges in G with both endpoints in $O_1$, normalized by the number of vertices in $O_1$. The same holds for any matrix derived similarly from a refinement of the partition $\text{StabOrb}_G(v)$.*

### 3.3.2 The Eigenvalues of $G_k$

We define a partition of the vertex set of $G_k = (V, E)$ into 5 classes (for $k \geq 3$) and prove that it forms a refinement of $\text{StabOrb}_{G_k}(v_0)$, where $v_0$ denotes the vertex $(e_1, e_1)$. The classes of the partition are defined as follows.

$$
\begin{aligned}
V_1 &= \{v_0\}, \\
V_2 &= \{(v_1, v_2) \in V \mid (v_1)_1 = (v_2)_1 = 0\}, \\
V_3 &= \{(v_1, v_2) \in V \mid \{(v_1)_1, (v_2)_1\} = \{0, 1\}\}, \\
V_4 &= \{(v_1, v_2) \in V \mid e_1 \in \{v_1, v_2\}, v_1 \neq v_2\}, \\
V_5 &= \{(v_1, v_2) \in V \mid (v_1)_1 = (v_2)_1 = 1, e_1 \notin \{v_1, v_2\}\}.
\end{aligned}
$$

It is not difficult to show that $\{V_i\}_{i=1}^{5}$ is a partition of $V$ and that $|V_1| = 1$, $|V_2| = 2^{k-2} \cdot (2^{k-1} - 1)$, $|V_3| = 2^{k-1} \cdot (2^{k-1} - 1)$, $|V_4| = 2 \cdot (2^{k-1} - 1)$, and $|V_5| = (2^{k-2} - 1) \cdot (2^{k-1} - 1)$.

**Lemma 3.18.** *For every $k \geq 3$, the partition $\{V_i\}_{i=1}^{5}$ of $V$ is a refinement of the orbit set $\text{StabOrb}_{G_k}(v_0)$.*

**Remark 3.19.** *In fact, it turns out the partition $\{V_i\}_{i=1}^{5}$ equals $\text{StabOrb}_{G_k}(v_0)$ for any $k \geq 3$ (and not a strict refinement of it), but we do not need it in our proof.*

**Proof:** We need to prove that for every two vertices $u, v$ in the same vertex class $V_i$ (for some $1 \leq i \leq 5$) there exists an $f \in \text{Aut}(G_k)$ such that $f(u) = v$ and $f(v_0) = v_0$. For each $i$ (except for the trivial case of $i = 1$), the required automorphism will maintain the first bit of both vectors in a given vertex. On the remaining bits the automorphism can be defined similarly to the automorphism in the proof of vertex-transitivity of $G_k$ (Lemma 3.6). The case of $i = 5$ is slightly more involved, and we give the details below.

Observe that it suffices to prove the existence of $f$ as above for $u = ((1, e_1), (1, e_2))$ where $e_1$ and $e_2$ are of length $k - 1$. Denote $v = (v_1, v_2)$, and let $v'_1$ and $v'_2$ denote the last $k - 1$ coordinates of the vectors $v_1$ and $v_2$ respectively. Notice that both $v'_1$ and $v'_2$ are nonzero and that $\langle v'_1, v'_2 \rangle = 0$. We define a $(k - 1) \times (k - 1)$ matrix $B$ as follows: the first column is $v'_1$, the second column is a certain vector that has inner product 1 with $v'_2$ (such a vector exists since $v'_2$ is nonzero), and the remaining columns are vectors which, together with $v'_1$, form a basis of the space of vectors orthogonal to $v'_2$ (such vectors exist since $v'_1$ is nonzero and is orthogonal to $v'_2$). Now, define a $k \times k$ matrix $A$ that has $e_1$ (of length $k$) as its first row and column, and the remaining entries consist of the matrix $B$. Observe that $A$ is invertible and that the function $f$ that maps any vertex $(x, y) \in V_5$ to $(Ax, A^{-t}y)$ is an automorphism of $G_k$. In addition, $f$ satisfies $f(e_1, e_1) = (Ae_1, A^{-t}e_1) = (e_1, e_1)$ and $f(u) = ((1, Be_1), (1, B^{-t}e_2)) = ((1, v'_1), (1, v'_2)) = (v_1, v_2)$. ∎

Now, a straightforward calculation shows that the matrix defined in Corollary 3.17 is

$$
M_k = \begin{pmatrix}
0 & 2^{k-2} \cdot (2^{k-1} - 1) & 0 & 0 & 0 \\
1 & 2^{k-3} \cdot (2^{k-2} - 1) & 2^{k-2} \cdot (2^{k-2} - 1) & 2 \cdot (2^{k-2} - 1) & (2^{k-2} - 1) \cdot (2^{k-3} - 1) \\
0 & 2^{k-3} \cdot (2^{k-2} - 1) & 2^{k-2} \cdot (2^{k-2} - 1) & 2^{k-2} & 2^{k-3} \cdot (2^{k-2} - 1) \\
0 & 2^{k-2} \cdot (2^{k-2} - 1) & 2^{2k-4} & 0 & 0 \\
0 & 2^{k-2} \cdot (2^{k-3} - 1) & 2^{2k-4} & 0 & 2^{2k-5}
\end{pmatrix}.
$$

It can be verified that the the eigenvalues of the matrix $M_k$ are

$$2^{k-2} \cdot (2^{k-1} - 1), \ 2^{\frac{3k}{2}-3}, \ 2^{k-3}, \ -2^{k-2}, \ -2^{\frac{3k}{2}-3}.$$

Therefore, these are the eigenvalues of our graph $G_k$ as well, and in particular its smallest eigenvalue is $-2^{\frac{3k}{2}-3}$, as required for Lemma 3.8.

# 4 Algorithms for Linear Index Coding

Let us start with some useful results related to graph coloring.

## 4.1 Graph Coloring

We recall the following simple claim (simplified and rephrased from [7]), which reduces the algorithmic task of coloring a graph to the algorithmic task of finding a large independent set in it.

**Claim 4.1.** *Let $\mathcal{G}$ be a graph family which is closed under taking induced subgraphs, let $c_1, c_2 > 1$ be arbitrary fixed constants, and let $f : \mathbb{N} \to \mathbb{N}$ be any non-decreasing function satisfying $c_1 f(n) \leq f(2n) \leq c_2 f(n)$ for all sufficiently large $n$. Then if there exists a (randomized) polynomial time algorithm which finds an independent set of size $f(n)$ in any $n$ vertex graph $G \in \mathcal{G}$, then there exists a (randomized) polynomial time algorithm which finds an $O(\frac{n}{f(n)})$-coloring of any $n$ vertex graph $G \in \mathcal{G}$.*

Note that the condition on the function $f$ above is satisfied, for example, by any function defined by $f(n) = n^\alpha \log^\beta n$ for some fixed constants $0 < \alpha < 1$ and $\beta > 0$.

A major component in our algorithms is the following theorem of Karger, Motwani and Sudan [21].

**Theorem 4.2** ([21]). *There exists a randomized polynomial time algorithm that given an $n$ vertex graph $G$ with maximum degree at most $\Delta$ and $\chi_v(G) \leq \kappa$ for some $\kappa \geq 2$, finds an independent set of size $\widetilde{\Omega}(\frac{n}{\Delta^{1-2/\kappa}})$.*

## 4.2 An Algorithm for Linear Index Coding for Graphs with Constant Minrank

In this section we present our algorithm for linear index coding for graphs with minrank at most $k$ and analyze it. For every $k$, denote $\kappa_k = 2^{\frac{k}{2}} + 1 - 2^{1-\frac{k}{2}}$. Recall that, by Theorem 3.10, every graph $G$ whose complement has minrank at most $k$ satisfies $\chi_v(G) \leq \kappa_k$. Our main result is the following.

**Theorem 4.3.** *There exists a randomized polynomial time algorithm that given an $n$ vertex graph $G$ satisfying $\mathrm{minrk}_2(G) \leq k$, finds a linear index code for $G$ of length $\widetilde{O}(n^{1-g(k)})$, where $g : \mathbb{N} \to (0, 1]$ is the function defined by $g(1) = g(2) = 1$ and for any $k \geq 3$,*

$$g(k) = \frac{g(k-1)}{g(k-1) + 1 - \frac{2}{\kappa_k}}.$$

> **Minrank-Basic**(G, k)
> Input: A graph $G$ with $\text{minrk}_2(\overline{G}) \leq k$.
>
> - If $k \in \{1, 2\}$ return the larger color class in a 2-coloring of $G$.
>
> - Otherwise, return the larger independent set between the following two:
>
>   - The independent set found by the Karger-Motwani-Sudan algorithm (Theorem 4.2) applied to $G$.
>   - The independent set obtained by running Minrank-Basic$(G[\Gamma(v)], k-1)$, where $v$ is a vertex of maximum degree.

Figure 1: Algorithm **Minrank-Basic**

As mentioned earlier, in order to find a linear index code for a graph $G$ it suffices to find a coloring of its complement. In order to simplify the notations, we consider the input graph $G$ as a graph whose *complement* has minrank at most $k$ and the goal is to color $G$. Theorem 4.3 stems from combining Claim 4.1 and the following theorem.

**Theorem 4.4.** *There exists a randomized polynomial time algorithm that given an n vertex graph G satisfying $\text{minrk}_2(\overline{G}) \leq k$, finds an independent set of size $\widetilde{\Omega}(n^{g(k)})$, where g is the function defined in Theorem 4.3.*

**Proof:** The proof is by induction on $k$. For $k \in \{1, 2\}$ the graph is 2-colorable and thus it is easy to efficiently find an independent set of size at least $\frac{n}{2}$ in $G$.

Assume that $k \geq 2$ and let $\Delta$ denote the maximum degree in $G$. Recall that the assumption $\text{minrk}_2(\overline{G}) \leq k$ implies that $\chi_v(G) \leq \kappa_k$. We consider two ways of finding an independent set in $G$ (see Figure 1). The first is by applying the Karger-Motwani-Sudan algorithm to $G$, which by Theorem 4.2 finds an independent set of size $\widetilde{\Omega}\left(\frac{n}{\Delta^{1-2/\kappa_k}}\right)$. The second is by applying the induction hypothesis to the subgraph $G[\Gamma_G(v)]$ of $G$ induced by the neighbors of a vertex $v$ of degree $\Delta$. By Lemma 2.8, $\text{minrk}_2(\overline{G[\Gamma_G(v)]}) \leq k-1$, and thus it is possible to find an independent set in $G[\Gamma_G(v)]$, which is also an independent set in $G$, of size $\widetilde{\Omega}(\Delta^{g(k-1)})$. This implies that at least one of the two independent sets has size

$$\widetilde{\Omega}\left(\max\left(\frac{n}{\Delta^{1-2/\kappa_k}}, \Delta^{g(k-1)}\right)\right) = \widetilde{\Omega}(n^{g(k)}),$$

where the equality follows from the definition of $g$. ∎

We note that our algorithm is essentially identical to the algorithm of [21]. However, the analysis of the second way of finding an independent set is different. Recall that here the algorithm is applied recursively to the graph $G[\Gamma(v)]$ where $v$ is a vertex of maximum degree. In [21], the analysis uses Lemma 2.3, that says that if $\chi_v(G) \leq \kappa$ then $\chi_v(G[\Gamma(v)]) \leq \kappa - 1$. However, in our algorithm, which colors a graph $G$ whose complement has minrank at most $k$, we are able to use

Lemma 2.8, which says that, for such graphs, $\overline{G[\Gamma(v)]}$ has minrank at most $k-1$. This yields that $G[\Gamma(v)]$ has vector chromatic number at most $\kappa_{k-1}$. It is not difficult to see that for every $k \geq 3$ it holds that $\kappa_{k-1} < \kappa_k - 1$, and therefore our analysis improves upon the analysis of [21] for our graphs. For example, for $k = 3$ our analysis shows that the algorithm finds an independent set of size $\tilde{\Omega}(n^{0.7357})$ since $g(3) = \frac{3+\sqrt{2}}{6} \approx 0.7357$, whereas the analysis of [21] guarantees an independent set of size $\tilde{\Omega}(n^{0.7247})$.

### 4.3 An Improved Algorithm for Linear Index Coding for Graphs with Minrank 3

In this section we adapt recent improvements in algorithms for coloring 3-colorable graphs and their analysis by Arora, Chlamtac and Charikar [4, 11]. This improvement arises from a more refined analysis of the rounding algorithm of a semidefinite program of [21], as well as a new rounding algorithm which exploits this analysis. It turns out that these improvements can be used for graphs with arbitrary (strict) vector chromatic number $\kappa$, and not only for 3-colorable graphs. This allows us to apply them to graphs $G$ with $\text{minrk}_2(\overline{G}) = k$ for constant $k$. We demonstrate this approach for the case of $k = 3$, but we note that this can be extended to any $k \geq 4$ as well.

Recall that every graph whose complement has minrank $k$ has (strict) vector chromatic number at most $\kappa_k$. For $k = 3$ we have $\kappa_3 \approx 3.1213$. The following theorem summarizes the guarantee that arises from applying the aforementioned improvements to our setting (for comparison, see Theorem 4.2 for $\kappa = \kappa_3$). While the analysis is somewhat technical, it follows the same lines as the analysis in [4]. For the sake of completeness, we include the proof in Section 4.4.

**Theorem 4.5.** *There exists a randomized polynomial time algorithm that given an n vertex graph $G$ satisfying $\text{minrk}_2(\overline{G}) \leq 3$, and has maximum degree at most $\Delta = n^{0.7426}$, finds an independent set of size $\tilde{\Omega}(n\Delta^{-0.96452(1-2/\kappa_3)}) \geq \tilde{\Omega}(n^{0.7426})$.*

Theorem 1.2 follows easily from the following corollary, using Claim 4.1.

**Corollary 4.6.** *There exists a randomized polynomial time algorithm that given an n vertex graph $G$ satisfying $\text{minrk}_2(\overline{G}) \leq 3$, finds an independent set of size $\tilde{\Omega}(n^{0.7426})$.*

**Proof:** The proof resembles that of Theorem 4.4. Suppose that there exists a vertex $v \in V$ with degree at least $\Delta = n^{0.7426}$. By Lemma 2.8, $\overline{G[\Gamma(v)]}$ has minrank at most 2, and hence $G[\Gamma(v)]$ is 2-colorable. Thus, we can find an independent set of size at least $\frac{\Delta}{2}$ by taking one of the two color classes. Otherwise, the maximum degree is at most $\Delta$, and the corollary follows by Theorem 4.5. ∎

### 4.4 Proof of Theorem 4.5

Recall that the density function of the *standard normal distribution* is the function that maps any $t \in \mathbb{R}$ to $\frac{1}{\sqrt{2\pi}}e^{-t^2/2}$. A random vector $\zeta = (\zeta_1, \ldots, \zeta_n)$ is said to has the *n-dimensional standard normal distribution* if the components $\zeta_i$ are independent and each has the standard normal distribution. Note that this distribution is invariant under rotation, and its projections onto orthogonal subspaces are independent. In particular, for any unit vector $w \in \mathbb{R}^n$, the projection $\langle \zeta, w \rangle$ has the standard normal distribution. We denote the corresponding tail bound by

$$N(s) = \int_s^\infty \frac{1}{\sqrt{2\pi}} e^{-\frac{t^2}{2}} dt.$$

The following property of the normal distribution ([15], Chapter VII) will be used.

**Lemma 4.7.** *For any $s > 0$, $\frac{1}{\sqrt{2\pi}} \left( \frac{1}{s} - \frac{1}{s^3} \right) e^{-s^2/2} \leq N(s) \leq \frac{1}{\sqrt{2\pi}s} e^{-s^2/2}$.*

Consider the following slight variant of the Karger-Motwani-Sudan algorithm proposed in [4].

---

**KMS**$'(G = (V, E), (w_i)_{i \in V})$

- For all $t > 0$,
    - Choose $\zeta \in \mathbb{R}^n$ from the $n$-dimensional standard normal distribution, and let $V_\zeta(t) = \{i \in V \mid \langle \zeta, w_i \rangle \geq t\}$.
    - Pick any edge $\{i, j\} \in E$ with both endpoints in $V_\zeta(t)$, and eliminate both $i$ and $j$. Repeat until no such edges are left.
    - Let $V'_\zeta(t)$ be the set of remaining vertices in $V_\zeta(t)$.
- For $t$ which maximizes $\left| V'_\zeta(t) \right|$, return the independent set $V'_\zeta(t)$.

---

Figure 2: Algorithm **KMS**$'$

The following lemma summarizes the main technical result of [21]. It is easily seen that Theorem 4.2 that was used in Section 4 follows directly from this lemma.

**Lemma 4.8.** *For every constant $\sigma > 0$, and every graph $G = (V, E)$ with maximum degree at most $\Delta$, there exists a threshold $t = t(\Delta)$ such that $N(t) = \tilde{\Omega}(\Delta^{-(1-\sigma)/(1+\sigma)})$ and such that if $(w_i)_{i \in V}$ is a strict vector $(1 + 1/\sigma)$-coloring of $G$ (edges correspond to inner product $-\sigma$), then in algorithm KMS$'(G, (w_i)_{i \in V})$ we have for every vertex $i$,*

$$\Pr_\zeta[i \in V'_\zeta(t) \mid i \in V_\zeta(t)] \geq \frac{1}{2}. \tag{3}$$

Lemma 4.8 shows that, in algorithm KMS$'$, it is sufficient to choose the threshold $t$ such that $N(t) = \tilde{\Theta}(\Delta^{-(1-\sigma)/(1+\sigma)})$ in order to guarantee that the probability in (3) is large for every vertex. Note that in order to find an independent set of size $\Omega(nN(t))$, it suffices to guarantee that this probability will be large for at least half the vertices. The improved analysis of Arora et al. [4, 11] shows that under certain conditions, this must happen, even for slightly smaller values of $t$ (i.e., larger values of $N(t)$). This, in turn, guarantees a larger independent set. Specifically, they show the following (rephrased here for general values of $\chi_v^{(s)}(G)$).

**Theorem 4.9** ([11], Theorem 3.4.5). *For every graph $G$ with strict vector $(1 + 1/\sigma)$-coloring $(w_i)_{i \in V}$ and maximum degree at most $\Delta$, either*

- Inequality (3) holds for at least half the vertices, or,

- letting $c > 0$ be such that $N(t) = \Delta^{-(1-\sigma)/[(1+\sigma)(1+c)]}$, there exists a vertex $i \in V$, a constant $\alpha \in [0, \frac{c}{1+c}]$ and a set of vertices $W \subseteq V$ such that

    1. For all $j \in W$,
    $$\left|\langle w_i, w_j \rangle - \mu_\sigma(\alpha)\right| = O(1/\log n), \tag{4}$$
    where $\mu_\sigma$ is defined by $\mu_\sigma(\alpha) = \sigma^2 + (1-\sigma^2)\alpha$.

    2. Letting $z_j = w_j - \langle w_i, w_j \rangle w_i$ (the component of $w_j$ orthogonal to $w_i$), we have
    $$\Pr{}_\zeta[\exists j \in W : \langle \zeta, z_j \rangle \geq \rho_\sigma(c,\alpha)(1+\sigma)t - o(t)] = \Omega(1/\log n), \tag{5}$$
    where $\rho_\sigma$ is the function defined by
    $$\rho_\sigma(c,\alpha) = 1 + \sigma - \sigma\alpha - \sqrt{(1-\alpha^2)c}.$$

**Remark 4.10.** *Equation (4) implies that for all $j \in W$,*
$$\left|\|z_j\|^2 - (1-\sigma^2)\pi_\sigma(\alpha)\right| = O(1/\log n), \tag{6}$$
*where $\pi_\sigma$ is defined by*
$$\pi_\sigma(\alpha) = (1-\alpha)(1+\sigma^2 + \alpha(1-\sigma^2)).$$

**Remark 4.11.** *The set $W$ in Theorem 4.9 is actually a subset of the neighbors of neighbors of $i$, $\Gamma(\Gamma(i))$, however this is not used anywhere in the analysis.*

**Lemma 4.12.** *Let $G, \sigma, (w_i)_{i \in V}, \Delta$ be as in Theorem 4.9, and write $\Delta = n^\delta$. Then if a constant $c > 0$ satisfies*
$$\min_{0 \leq \alpha \leq c/(1+c)} \left( \frac{(\rho_\sigma(c,\alpha))^2}{\pi_\sigma(\alpha)(1+c)} + \phi_\sigma(c,\alpha) \right) > \frac{1}{\delta}, \tag{7}$$
*where $\phi_\sigma$ and $s_\sigma$ are defined by*
$$\phi_\sigma(c,\alpha) = \frac{1-\sigma}{(1+\sigma)(1+c)} - \frac{1-s_\sigma(\alpha)}{1+s_\sigma(\alpha)}$$
*and*
$$s_\sigma(\alpha) = \frac{\sigma + (\mu_\sigma(\alpha))^2}{1 - (\mu_\sigma(\alpha))^2},$$
*then there exists a threshold $t$ such that $N(t) = \tilde{\Theta}\left(\Delta^{-(1-\sigma)/((1+\sigma)(1+c))}\right)$ and either*

1. *$\text{KMS}'(G, (w_i)_{i \in V})$ returns an independent set of size $\Omega(nN(t)) = \tilde{\Omega}\left(n\Delta^{-(1-\sigma)/((1+\sigma)(1+c))}\right)$, or*

2. *there exists a set $W$ as in Theorem 4.9 for which algorithm $\text{KMS}'(G[W], (z_j)_{j \in W})$ returns an independent set of size $\Omega(nN(t))$.*

**Proof:** Assume that case 1 does not hold. That is, for the above $c$ and $t$, KMS' does not return an independent set of size $\Omega(nN(t))$. Then, by Theorem 4.9, the set $W$ as described in the theorem must exist. Since, by (4) the vectors $(w_j)_{j \in W}$ all have a common component of length at least $\mu_\sigma - o(1)$, the vectors $(z_j)_{j \in W}$ constitute a vector $(1 + 1/(s_\sigma - o(1)))$-coloring of $G[W]$. Thus, by Theorem 4.2, we can find an independent set in $G[W]$ of cardinality $\tilde{\Omega}(|W|\Delta^{-(1-s_\sigma(\alpha))/(1+s_\sigma(\alpha))-o(1)})$. Thus, for $|W| \geq n\Delta^{-\phi_\sigma(c,\alpha)+o(1)}$, case 2 above follows.

Assume this is not the case. That is, $|W| \leq n\Delta^{-\phi_\sigma(c,\alpha)+o(1)}$. By a union bound (and using (6)), the probability of the event in (5) is at most

$$|W|N\left(\left(\sqrt{\frac{1+\sigma}{1-\sigma}} \cdot \frac{\rho_\sigma(c,\alpha)}{\sqrt{\pi_\sigma(\alpha)}} - o(1)\right)t\right).$$

Moreover, by Lemma 4.7, for any constant $\beta > 1$ we have $N(\beta t) \leq \tilde{O}(N(t)^{\beta^2})$. Applying this bound, using our chosen value of $N(t)$, the above is at most

$$\tilde{O}\left(|W|\Delta^{-(\rho_\sigma(c,\alpha))^2/(\pi_\sigma(\alpha)(1+c))+o(1)}\right),$$

which given our bound on $|W|$ and our value of $c$ contradicts (5), meaning that case 1 must hold. ∎

Theorem 4.5 now follows as an immediate corollary of the above lemma (for $c = 0.03678$ and $\delta = 0.7426$). The algorithm below gives the corresponding guarantee.

---

**Augmented-KMS**($G$)

- Compute the optimum strict vector coloring $(w_i)_{i \in V}$ of $G$.

- For every vertex $i$:

    - For every $b \in (-1, 1)$, let $W_i(b) = \{j \in V \mid 1 > \langle w_i, w_j \rangle \geq b\}$ and for all $j \in V$ let $z_j = w_j - \langle w_i, w_j \rangle w_i$. For all $b$, run KMS'($G[W_i(b)], (z_j/\|z_j\|)_{j \in W_i(b)}$).

- Run KMS'($G, (w_i)_{i \in V}$).

- Return the largest of the independent sets found above.

---

Figure 3: Algorithm **Augmented-KMS**

## 5 Discussion

The approach taken in this paper for finding a short linear index code for a given graph relies on coloring the complement of the graph. This is done via studying the maximum vector chromatic number of graphs whose complement has minrank $k$. Recall that $G_k$ is the graph that maximizes both the chromatic number and the vector chromatic number among the graphs whose complement has minrank $k$. This graph satisfies

$$k = \operatorname{minrk}_2(\overline{G_k}) < \chi_v(G_k) = \chi_v^{(s)}(G_k) < \chi(G_k),$$

which implies that we have a much stronger guarantee for the vector chromatic number than for the chromatic number. Indeed, while by Corollary 3.9 we have $\chi_v(G_k) \leq 2^{\frac{k}{2}} + 1$, Alon [2] has recently shown that $\chi(G_k) \geq 2^{k-O(\sqrt{k \log k})} = 2^{(1-o(1))k}$.

For graphs with bounded chromatic number (as opposed to bounded vector chromatic number), stronger results exist, such as the combinatorial algorithm of Blum [7] and the SDP hierarchy approach of Chlamtac [10]. However, it seems difficult to gain any additional improvements from such techniques, as the worst-case bound on the chromatic number is strictly weaker than the bound on the vector chromatic number.

A possible interesting alternative would be to rely directly on the minrank guarantee, and thus get around the gap between the minrank and the chromatic number of the complement. One natural SDP relaxation for minrank relies on the characterization of graphs with minrank at most $k$ as graphs whose complements have a homomorphism to $G_k$ (see Section 3). This leads to a natural relaxation, similar to relaxations for the Unique Games problem (see, e.g., [9]):

$$
\begin{aligned}
\min \quad & k \\
\text{s.t.} \quad & \exists \{v_{x,i} \mid x \in V(G), i \in V(G_k)\} \\
\text{s.t.} \quad & \langle v_{x,i}, v_{x,j} \rangle = 0 && \forall x \forall i \neq j \\
& \langle v_{x,i}, v_{y,j} \rangle = 0 && \forall x, y, i, j \text{ s.t. } \{x,y\} \notin E(G) \text{ and } (i = j \text{ or } \{i,j\} \notin E(G_k)) \\
& \sum_{i,j \in V(G_k)} \langle v_{x,i}, v_{y,j} \rangle = 1 && \forall x, y \in V(G) \text{ (including } x = y\text{)}
\end{aligned}
$$

It turns out that there is a rounding for the above SDP relaxation for minrank which matches our performance guarantee. Indeed, we have the following.

**Claim 5.1.** *Let $\{v_{x,i}\}$ be a solution to the above SDP relaxation with value $k$, and let $\{u_i\}$ be an optimum strict vector coloring for $G_k$. Then the vectors $\{w_x \mid x \in V(G)\}$ defined by*

$$w_x = \sum_{i \in V(G_k)} v_{x,i} \otimes u_i$$

*are a vector $\kappa$-coloring of $\overline{G}$, where $\kappa = \chi_v(G_k)$.*

It remains open whether one can achieve an improved rounding for this relaxation (or possibly stronger relaxations).

It would also be interesting to extend our algorithms to the case of *general* (non-linear) index coding. We note that it was shown in [26] that the complement of a graph that has an index code of length $k$ has a coloring that uses at most $2^{2^k}$ colors. This, together with the coloring algorithm of [20], immediately implies an algorithm for general index coding. However, it is possible that properties of graphs with bounded length index codes can be exploited to improve the guarantee of this algorithm. The techniques used in this paper, though, seem to be beneficial only for linear index coding which is characterized by the minrank parameter.

## Acknowledgement

We would like to thank Noga Alon and Michael Langberg for fruitful discussions.